\documentclass[reprint,
superscriptaddress,
amsmath,amssymb,aps,showkeys,showpacs,
twoside,final,secnumarabic,%raggedbottom,
nofootinbib]{revtex4-1}

\usepackage[paperwidth=205mm,paperheight=290mm,top=17mm,bottom=25mm,
inner=17mm,outer=17mm,
twoside]{geometry}

\usepackage{cmap} % Улучшенный поиск %русских слов в~полученном pdf-файле
\usepackage[T1,T2A]{fontenc}
\usepackage[utf8]{inputenc}
\usepackage[russian,english]{babel}
\usepackage{color}
\usepackage{graphicx}% Include figure files
\usepackage{dcolumn}% Align table columns on decimal point
\usepackage{bm} % bold math
\usepackage[unicode=true,colorlinks=true,linkcolor=magenta, urlcolor=blue, citecolor = blue,breaklinks]{hyperref}
\usepackage{multirow}
\usepackage{url}
\usepackage{breakurl}
\DeclareGraphicsExtensions{.eps}

\newcount\issue
\newcount\Vol
\newcount\numb
\usepackage{fancyhdr} %this packages %provides fancy up and bottom of page
\def\Vol{\textbf{78}}
\def\numb{x}
\begin{document}

%====== Начало шапки статьи  ============
\title{JOURNAL SECTION OR CONFERENCE SECTION\\[20pt]
An interstellar mission to the closest black hole?} 

\def\addressa{Center for Astronomy and Astrophysics, Center for Field Theory and Particle Physics, and Department of Physics, Fudan University, Shanghai 200438, China}
\def\addressb{School of Humanities and Natural Sciences, New Uzbekistan University, Tashkent 100001, Uzbekistan}

\author{\firstname{Cosimo}~\surname{Bambi}}
\email[E-mail: ]{bambi@fudan.edu.cn }
\affiliation{\addressa}
\affiliation{\addressb}

\received{xx.xx.2025}
\revised{xx.xx.2025}
\accepted{xx.xx.2025}

\begin{abstract}
In this manuscript, I discuss the possibility of sending a small probe to the closest black hole with the goal of addressing some fundamental questions of modern physics. Are astrophysical black holes the Kerr black holes predicted by General Relativity? Do astrophysical black holes have an event horizon? Is the physics around a black hole the same physics as in our laboratories on Earth? While we do not have the technology for a similar mission today, it may be available in the next 20-30~years. The whole mission may last up to a century (depending on the actual distance of the black hole and the speed of the probe), but it may represent a unique opportunity to perform precise and accurate tests of General Relativity in the strong field regime. 
\end{abstract}

\par
\keywords{Black Holes; Interstellar Missions; Laser Propulsion; Tests of General Relativity   \\[5pt]}
%DOI:  

\maketitle
\thispagestyle{fancy}

%====== Начало  статьи  ============

\section{Introduction}

Stellar-mass black holes are the final products of the evolution of heavy stars, those with masses exceeding $\sim 20$~$M_\odot$~\cite{Bambi:2025rod}. Current astrophysical models predict that in our Galaxy there are $10^8$-$10^9$~black holes formed from the collapse of heavy stars~\cite{Olejak:2019pln,Timmes:1995kp} and simple estimates suggest that the closest black hole may be at 20-40~light years from the Solar System (see next section). We can thus wonder whether it could be possible so send a probe to the closest black hole with the goal of testing the actual nature of that object and addressing some fundamental questions of modern physics~\cite{Bambi:2025kcr}:
\begin{enumerate}
\item Are astrophysical black holes the Kerr black holes predicted by Einstein's theory of General Relativity? 
\item Do astrophysical black holes have an event horizon causally separating the black hole from the exterior region? 
\item Is the physics in the strong gravitational field of a black hole the same physics as in our laboratories on Earth?
\end{enumerate}

This idea was first discussed in Ref.~\cite{Bambi:2025kcr}. As I will show in the next sections, it is certainly very speculative and technologically challenging, but it is not completely unrealistic. Certainly it is impossible today, but it may be possible in the future. A key-point, which is not under our control, is the distance of the closest black hole. It is also clear that a similar mission can make sense only if it can do something that cannot be done with astrophysical observations from the Earth or with observatories orbiting the Earth~\cite{Bambi:2015kza,Bambi:2017khi,Bambi:2024kqz,Tripathi:2020yts,Bambi:2021chr}: this is not obvious {\it a priori} and should be accurately studied on the basis of the possible technology onboard the probe.

Studies of possible interstellar missions for explorations beyond the Solar System started at the end of the 1950s. Laser propulsion was proposed in the 1960s~\cite{p1,p2}. In the past 10-15~years, the exoplanet community has already discussed the possibility of sending small probes to study exoplanets in nearby stellar systems~\cite{lubin}. In this sense, the idea of sending a small probe to the closest black hole is not completely new and can be reformulated as the following question: if we can design an interstellar mission to study nearby exoplanets, can we do something similar to study black holes and test the predictions of General Relativity in the strong field regime?

\begin{figure*}[t]
\centering
\includegraphics[width=0.85\linewidth]{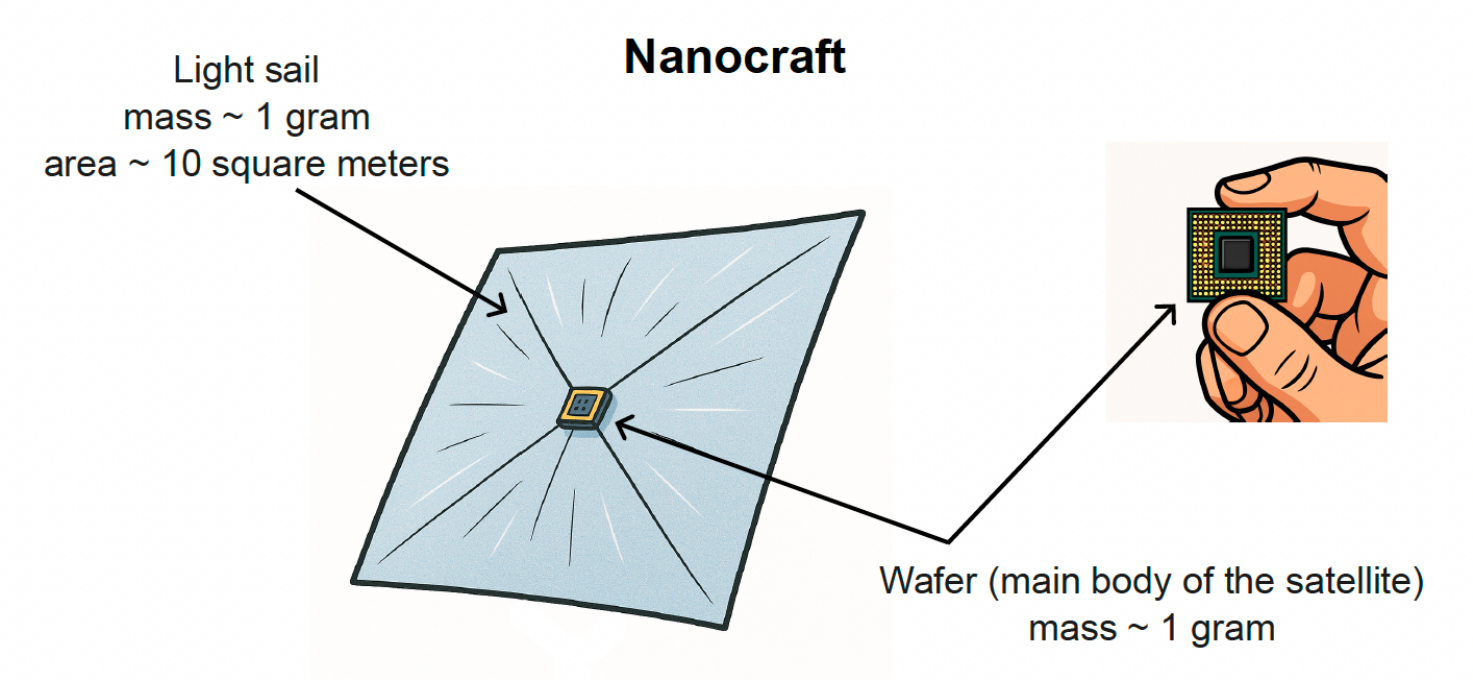}
%\vspace{-0.5cm}
\caption{Sketch of a nanocraft. The wafer is the main body of the satellite, with a computer processor, solar panels, navigation and communication equipment, etc. The light sail is necessary to accelerate the probe and can be used as an antenna for communication when the nanocraft is far from Earth. Figure from Ref.~\cite{Bambi:2025kcr}.}\label{f-nanocraft}
\end{figure*}

\section{The closest black hole}\label{sec:bh}

Let us estimate the distance of the closest black hole from the Solar System. The total mass in the form of stars in our Galaxy is estimated to be $M_{\rm stars} = ( 6.1 \pm 1.1 ) \cdot 10^{10}$~$M_\odot$~\cite{Licquia:2014rsa}. Most black holes in our Galaxy are isolated black holes, without any companion object~\cite{Olejak:2019pln}. The estimate of the total number of isolated black holes in our Galaxy, $N_{\rm BH}$, ranges from $\sim 1 \cdot 10^8$~\cite{Olejak:2019pln} to $\sim 1 \cdot 10^9$~\cite{Timmes:1995kp}. If we divide $M_{\rm stars}$ by $N_{\rm BH}$, we find that we expect one black hole per $\sim 600$~$M_\odot$ and $\sim 60$~$M_\odot$ of stellar mass, respectively. The stellar mass density in our region of the Galaxy is around 0.04~$M_\odot$~pc$^{-3}$~\cite{Lutsenko:2025aal}. The average volume per black hole around the Solar System should thus range from $\sim 15,000$~pc$^3$ (for $N_{\rm BH} = 10^8$) to $\sim 1,500$~pc$^3$ (for $N_{\rm BH} = 10^9$), which would correspond to one black hole in a sphere of radius $\sim 15$~pc and $\sim 7$~pc, respectively. If we expect one black hole within $\sim 7$~pc of the Solar System, the average distance between the Solar System and the black hole would be $\sim 5.5$~pc ($\sim 18$~light-years). For one black hole within $\sim 15$~pc of the Solar System, the average distance would be $\sim 12$~pc ($\sim 40$~light-years). In conclusion, from current astrophysical models we can estimate that the distance of the closest black hole from us may be 18-40~light-years.

While the closest black hole may be not very far from us, its detection can be very challenging. Isolated black holes can accrete from the interstellar medium and such an accretion process may produce a detectable flux of radiation~\cite{Shvartsman,Meszaros75,McDowell85,Campana93,Fujita:1997fh,Tsuna:2018abi,Kimura:2021ayq,Murchikova:2025oio}. Recently, Murchikova \& Sahu (2025)~\cite{Murchikova:2025oio} have shown that observational facilities like the Square Kilometer Array (SKA), the Atacama Large Millimiter/Submillimiter Array (ALMA), and James Webb Space Telescope (JWST) can detect isolated black holes in a warm interstellar medium within 150~light-years of Earth. The challenge is to identify these objects as accreting black holes. With a single telescope, we can just detect a faint source with a relatively featureless spectrum: such a source can be easily misclassified. We need to observe the full spectrum of a source to realize it is the spectrum of an isolated black hole accreting from the interstellar medium and this requires multi-telescope observations.

\section{The spacecraft}\label{sec:spacecraft}

Let us assume we discover a black hole at 20~light-years from Earth and we want to send a probe to study the object. It is clear that the probe must travel at a velocity of some fractions of the speed of light to reach the black hole within a reasonable time. This already rules out chemical propulsion of current rockets. The Tsiolkovsky rocket equation reads $m_i/m_f = e^{\Delta v/v_e}$, where $m_i$ is the initial total mass of the rocket (with propellant), $m_f$ is the final total mass of the rocket (without propellant), $\Delta v$ is the total change of the rocket's velocity, and $v_e$ is the effective exhaust velocity. If we use liquid hydrogen/liquid oxygen (such as in the American Space Shuttle and European Ariane launchers), $v_e \sim 4.5$~km/s: if we want the rocket to travel at 1/10 of the speed of light, even if $m_f$ were the proton mass, $m_i$ would largely exceed the total mass of the visible Universe.  

Today, laser propulsion~\cite{p1,p2} and nanocrafts~\cite{Lubin22,Kuhlmey25} appear to be the most promising solution for explorations beyond the Solar System. A nanocraft is a gram-scale spacecraft (see Fig.~\ref{f-nanocraft}). The main body of the probe is a gram-scale wafer: it includes a computer processor, solar panels, navigation and communication equipment, etc. The wafer is attached to an extremely thin, meter-scale light sail, which is necessary to accelerate the probe and can be used as an antenna for communication when the nanocraft is far from Earth. Ground-based high-power lasers can accelerate the nanocraft through the radiation pressure of their laser beams on the light sail. There are no specific technical problems to reach 90\% of the speed of light with this technique, but higher velocities increase significantly the total cost of the mission.

\section{The mission}\label{sec:mission}

The whole mission can be roughly divided into four phases: 1) acceleration of the probe, 2) interstellar trip, 3) approaching to the black hole and preparation for the scientific experiments, and 4) scientific experiments. 

If the black hole is at 20~light-years from Earth and we accelerate the nanocraft to reach 1/3~of the speed of light, the nanocraft can reach the black hole in 60~years, study the gravitational field of the black hole, and send all data to Earth, which can be received after 20~years. The total duration of the mission would be around 80~years. The distance of the black hole is not under our control and the duration of the mission increases as the distance of the black hole from the Solar System increases. The speed of the nanocraft is essentially determined by our laser technology and budget of the mission. For a given laser technology, if we increase the speed of the nanocraft we increase the costs of the mission. For a given speed of the nanocraft, a mora advanced technology can reduce the costs of the mission.

\subsection{Phase 1: acceleration of the probe}

In the acceleration phase, ground-based high-power lasers must be directed onto the nanocraft and the pressure of the photons in the laser beams onto the light sail should accelerate the nanocraft. The maximum acceleration that the nanocraft can endure is expected to be around $10^5$~m~s$^{-2}$~\cite{Kuhlmey25}. If we accelerate the nanocraft at the maximum acceleration, we need $10^3$~s ($\sim 17$~minutes) to reach 1/3 of the speed of light. In such a case, the acceleration distance would be around $5 \cdot 10^{10}$~m (about 1/3 of the distance Sun-Earth). When the nanocraft reaches the target velocity, the lasers are turned off, and the nanocraft can start its trip towards the black hole.  

The acceleration phase is certainly one of the critical parts of the mission, starting from its cost. With current technology, we could use an array of $10^8$ lasers, but its cost would be around one trillion EUR~\cite{Bandutunga}. This would be largely beyond the budget of any scientific experiment and makes such a mission impossible today. However, if we consider the trend of the price per coherent Watt in the past 20~years and we assume a similar trend for the next decades, we can expect that the cost of the lasers to accelerate the nanocraft will be around one billion EUR in 20-30~years, which is roughly the budget of a large experiment today.

The acceleration phase involves other critical parts. The nanocraft should be accelerated to the exact direction to reach the black hole, as later navigation can probably make only small corrections to the original trajectory of the nanocraft. The material and the properties of the light sail are crucial to guarantee a successful acceleration phase~\cite{Kuhlmey25}.

\begin{figure*}[t]
\centering
\includegraphics[width=0.40\linewidth]{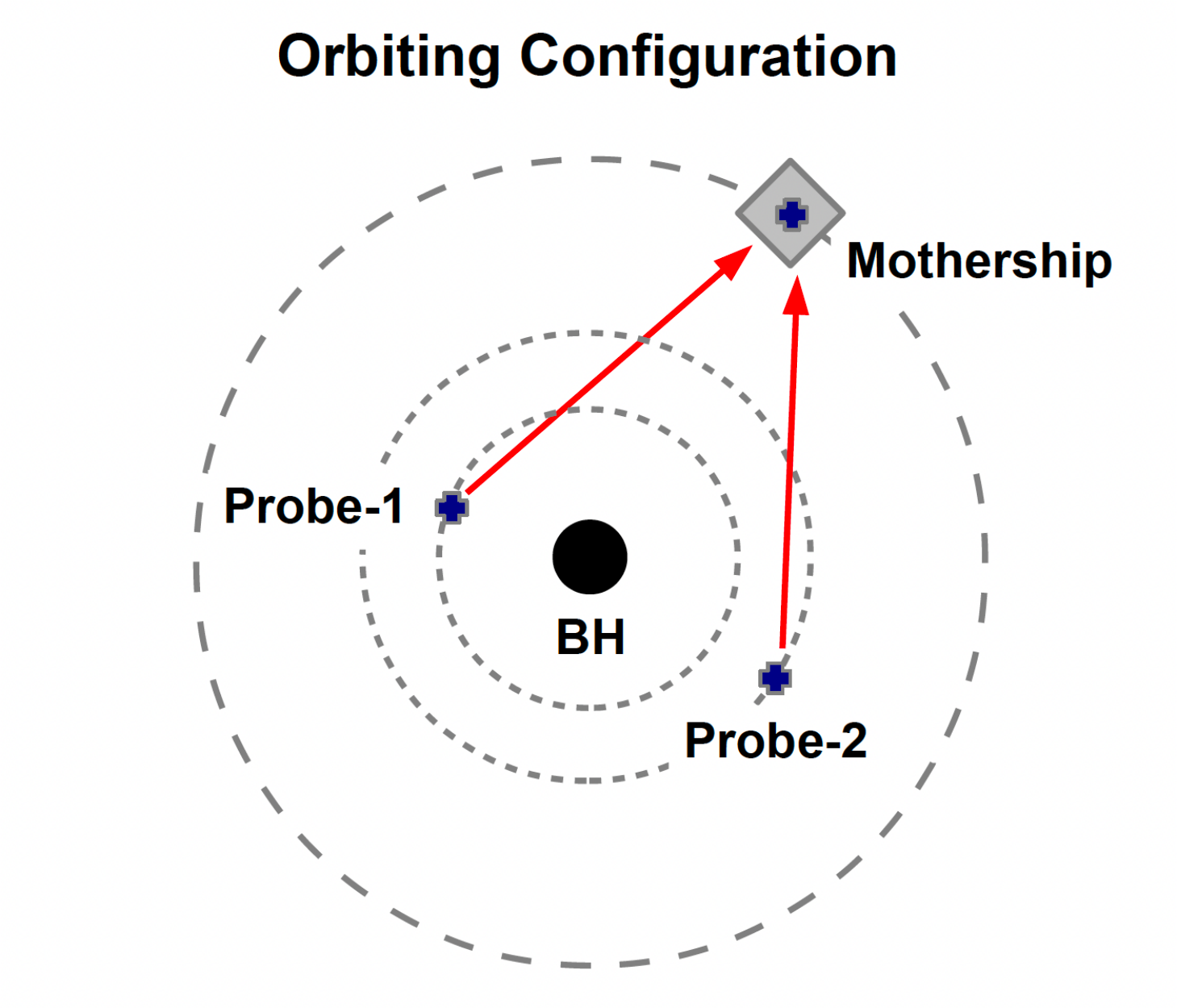}
\hspace{1.0cm}
\includegraphics[width=0.40\linewidth]{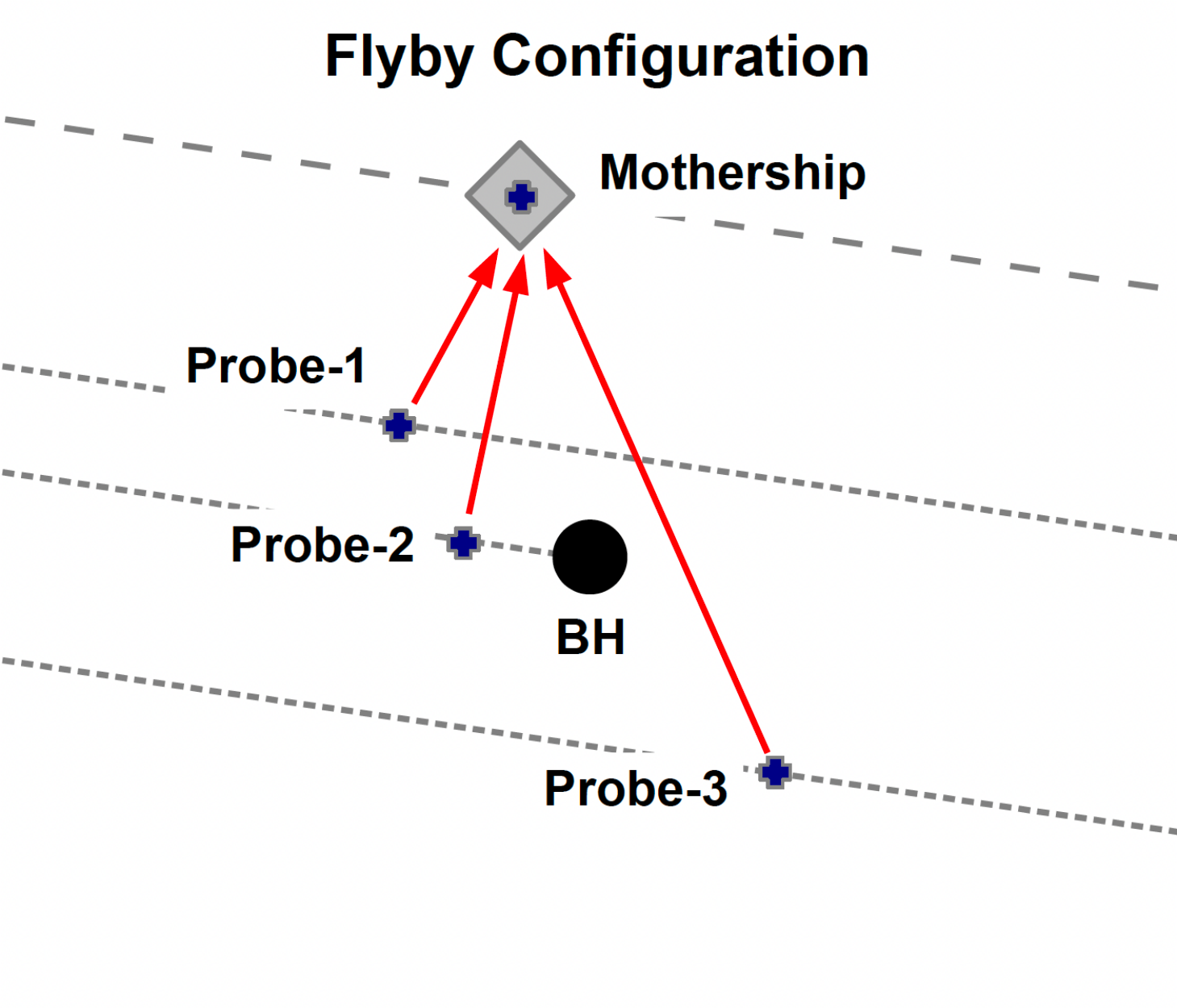}
%\vspace{-0.5cm}
\caption{{\bf Orbiting configuration (left picture):} the mothership with the light sail acting as an antenna to send the collected data to Earth orbits the black hole at a relatively large distance; the small probes orbit closer to the black hole and communicate with the mothership. {\bf Flyby configuration (right picture):} the spaceship fleet (mothership + small probes) passes through the gravitational field of the black hole; the mothership with the light sail acting as an antenna remains relatively far from the black hole and the small probes pass closer to the black hole; the small probes communicate with the mothership (and, if possible, among them). Some small probes (Probe-2 in the picture above) may fall onto the black hole.}\label{f-config}
\end{figure*}

\subsection{Phase 2: interstellar trip}

When the lasers are turned off, the nanocraft can start its trip towards the target black hole. Since the scientific experiments require two or more probes in the gravitational field of the black hole, after the acceleration phase the nanocraft may separate into one or more motherships and one or more smaller probes. In this way, even if one of the motherships/probes is damaged or lost during the long trip, the mission can continue and be successful.

\subsection{Phase 3: approaching to the black hole and preparation for the scientific experiments}

When the spaceship fleet is close to the black hole, it should get ready to start the scientific experiments in the gravitational field of the black hole. As shown in Fig.~\ref{f-config}, we may consider two configurations for the scientific experiments.

In the {\it orbiting configuration} (left picture in Fig.~\ref{f-config}), the mothership(s) and the small probe(s) orbit the black hole. In the picture in Fig.~\ref{f-config}, we have just one mothership and two small probes. The mothership has the light sail, which can be used as an antenna to send the collected data to Earth, and its orbit is relatively far from the black hole (if the mothership gets too close to the black hole, tidal effects could destroy the light sail). The small probes orbit closer to the black hole; they do not need any large antenna because the mothership is not too far from them. This configuration is certainly more suitable to test the gravitational field of the black hole, especially if the probes can complete many orbits around the black hole, because the precision of the measurements increases with the number of the orbits of the probes. However, it is not clear if and how the mothership and the probes can move from the unbound orbits of phase~2 to bound orbits around the black hole. Unbound circular orbits do not seem to be a solution, as they are unstable and cannot permit many orbits.  

In the {\it flyby configuration} (right picture in Fig.~\ref{f-config}), the spaceship fleet simply passes through the gravitational field of the black hole, which solves the problem of the change of the orbits. The challenge here is that the mothership(s) and the small probe(s) travel at high velocity and cross the strong gravity region in a very short time, so it is more challenging to perform very precise and accurate measurements of the strong gravitational field. A large number of small probes may help to perform more precise measurements in a shorter time.

\subsection{Phase 4: scientific experiments} 

The scientific experiments that can be done in the gravitational field of the black hole depend on the actual instrumentation and technology onboard the spaceship fleet. However, generally speaking we need to study how massive particles (the small probes) and electromagnetic signals (which are exchanged among the motherships and the small probes) propagate in the gravitational field of the black hole. In what follows, I will briefly discuss how we may address three fundamental questions. Are astrophysical black holes the Kerr black holes predicted by General Relativity? Do astrophysical black hole have an event horizon? Is the physics around a black hole the same physics as in our laboratories on Earth?

\subsubsection{Are astrophysical black holes the Kerr black holes predicted by General Relativity?}

In General Relativity and in the absence of exotic matter fields, the spacetime geometry around astrophysical black holes should be approximated very well by the Kerr solution. However, deviations from the Kerr metric are possible if General Relativity is not the correct theory of gravity, in the presence of exotic matter fields, and in scenarios with macroscopic quantum gravity effects. Tests of the spacetime metric around astrophysical black holes are usually referred to as tests of the {\it Kerr hypothesis}~\cite{Bambi:2015kza,Bambi:2017khi,Bambi:2024kqz,Tripathi:2020yts,Bambi:2021chr}. 

If the spacetime metric around a black hole is described by the Kerr solution, we can make very clear predictions about the trajectories of small particles in the gravitational field of the black hole and about the propagation of electromagnetic signals. This is exactly the situation of our spaceship fleet orbiting the black hole or passing through its gravitational field. If the small probes can communicate with the mothership(s), we can reconstruct their trajectories and how their electromagnetic signals propagate in the gravitational field of the black hole. The measurement of any deviation from the predictions of General Relativity could be caused by a violation of the Kerr hypothesis.

\subsubsection{Do astrophysical black hole have an event horizon?}

The {\it event horizon} is the boundary separating the black hole from the exterior region~\cite{Bambi:2017khi}. In General Relativity, if a particle falls onto a Kerr black hole and crosses its event horizon, an observer in the exterior region can only see that the particle approaches the black hole forever, without reaching the event horizon. If the particle sends a stable electromagnetic signal to the observer in the exterior region, the latter receives an electromagnetic signal that is more and more redshifted, without disappearing. In the case of our space mission, one of the probes can plunge onto the black hole and the mothership(s) and the other probes can measure the trajectory of the probe plunging onto the black hole and its electromagnetic signal. The frequency and the amplitude of such an electromagnetic signal will decrease, and at some point they will be so low and weak that the mothership(s) and the other probes will not be able to detect the signal any longer. This would be the prediction according to General Relativity.

There are scenarios beyond General Relativity where the final product of the gravitational collapse is not a black hole with an event horizon but some kind of compact horizonless object. An example is the fuzzball paradigm in string theory~\cite{Mathur:2024ify}, where instead of a black hole we have a fuzzball, which is a bound state of strings/branes without event horizon. When a particle hits the surface of the fuzzball, it should be converted into fuzzball degrees of freedom. In the case of our space mission, if one of the probes plunges onto a fuzzball, its electromagnetic signal sent to the mothership(s) and the other probes should instantly interrupt as soon as the probe hits the surface of the fuzzball. This may be clearly distinguished from the signal expected in the case of a Kerr black hole with an event horizon.

\subsubsection{Is the physics around a black hole the same physics as in our laboratories on Earth?}

In General Relativity and in any metric theory of gravity, the outcome of any local non-gravitational experiment is independent of where and when in the Universe it is performed~\cite{Will:2014kxa}. The physics in the strong gravitational field of a black hole is thus the same physics as in our laboratories on Earth. This is because in General Relativity and in any metric theory of gravity the gravitational field is described by the metric tensor $g_{\mu\nu}$ and we can always choose a locally inertial reference frame where the laws of physics are those of Special Relativity. However, there are models beyond General Relativity where this is not true. A common phenomenon in these models is the variation of fundamental constants~\cite{Uzan:2002vq,Uzan:2010pm,Bambi:2022lhq}: some of our fundamental constants (such as the fine structure constant $\alpha$, the electron mass $m_e$, etc.) may not be true constants of the ``fundamental'' theory and may depend on the configuration of the gravitational field. There are many studies reported in the literature on the possible temporal variation of fundamental constants and constraints from the study of distant sources. The case of possible variation of fundamental constants in strong gravitational fields is more difficult to study and currently there are no tests for gravitational fields stronger than those on the surface of white dwarfs~\cite{Berengut:2013dta}.

Let us assume we want to test the possible variation of the fine structure constant $\alpha$ in the strong gravitational field of the black hole. One of the probes could send the mothership two electromagnetic signals from fine structure doublets. Non-relativistic spectra depend mainly on the Rydberg constant ($\nu \propto R_\infty$). In the case of fine structure doublets, the frequency splitting between the two lines of the doublet is $\Delta\nu = \alpha^2 Z^4 R_\infty/ 2 n^3$, where $Z$ is the proton number of the nucleus and $n$ is the principal quantum number. If $\nu$ is the mean frequency of the doublet, $\Delta\nu/\nu \sim \alpha^2 Z^4$ and is independent of the photon redshift, so a variation of $\Delta\nu/\nu$ can be attributed to a variation of $\alpha$.

\section{Conclusions}\label{sec:conclusions}

In this manuscript, I outlined the idea of sending a small probe to the closest black hole with the goal of testing the actual nature of the object and address some fundamental questions of modern physics. The idea was first proposed in Ref.~\cite{Bambi:2025kcr}. While the idea is very speculative and technologically challenging, and certainly it is impossible today, it may be possible in the future.

The first step towards an interstellar mission to a black hole is certainly the discovery of a nearby black hole. This would permit us to have a specific target and estimate the actual chances of a similar mission. The distance of the closest black hole from us is not under our control but it is a crucial parameter for the feasibility of the mission. There are three possible situations:
\begin{enumerate}
\item If the black hole is within 20-25~light-years of Earth, the technology necessary for the mission may be developed in the next 20-30~years. The nanocraft may travel at 1/3~of the speed of light and the duration of the whole mission can be within a century.
\item If the black hole is not within 20-25~light-years of Earth, but still within 40-50~light-years, the technological requirements are more challenging. The probe should travel at a speed closer to the speed of light in order to remain with a mission with a duration within a century, and we may need more than 20-30~years to develop the technology necessary to make the mission possible. 
\item If the distance of the black hole is more than 40-50~light-years, it is not a problem of technology: the black hole is simply too far from us and, even if the nanocraft could travel close to the speed of light, the whole mission would last over a century. 
\end{enumerate}

Even in the most optimistic situation (case~1: the black hole is within 20-25~light-years), there are certainly many technological problems to solve. With the current laser technology, the mission would be simply too expensive. There are many other issues. How can we send the probe close to the black hole and how can we perform precise and accurate tests of the gravitational field of the black hole? How can the probe send the scientific data back to Earth? 

As of now, the idea of an interstellar mission to the closest black hole is only a speculative idea. However, it could become possible in the future.

\section*{FUNDING}
This work was supported by the National Natural Science Foundation of China (NSFC), Grant No.~W2531002, 12250610185, and 12261131497.

\section*{CONFLICT OF INTEREST}
The author declares that he has no conflicts of interest.

\nocite{*}

%%%%%%%%%%%%%%%%%%%%%%%%%%%%%%%%
% USE thebibliography
%%%%%%%%%%%%%%%%%%%%%%%%%%%%%%%%


\begin{thebibliography}{}

\bibitem{Bambi:2025rod}
C.~Bambi,
%``Stellar-Mass Black Holes,''
Symmetry \textbf{17}, 1393 (2025);
\href{https://doi.org/10.3390/sym17091393}{https://doi.org/10.3390/sym17091393}
[arXiv:2507.15270 [astro-ph.HE]].

\bibitem{Olejak:2019pln}
A.~Olejak, K.~Belczynski, T.~Bulik and M.~Sobolewska,
%``Synthetic catalog of black holes in the Milky Way,''
Astron. Astrophys. \textbf{638}, A94 (2020);
\href{https://doi.org/10.1051/0004-6361/201936557}{https://doi.org/10.1051/0004-6361/201936557}
[arXiv:1908.08775 [astro-ph.SR]].

\bibitem{Timmes:1995kp}
F.~X.~Timmes, S.~E.~Woosley and T.~A.~Weaver,
%``The Neutron star and black hole initial mass function,''
Astrophys. J. \textbf{457}, 834 (1996);
\href{https://doi.org/10.1086/176778}{https://doi.org/10.1086/176778}
[arXiv:astro-ph/9510136 [astro-ph]].

\bibitem{Bambi:2025kcr}
C.~Bambi,
%``An interstellar mission to test astrophysical black holes,''
iScience \textbf{28}, 113142 (2025);
\href{https://doi.org/10.1016/j.isci.2025.113142}{https://doi.org/10.1016/j.isci.2025.113142}
[arXiv:2504.14576 [gr-qc]].

\bibitem{Bambi:2015kza}
C.~Bambi,
%``Testing black hole candidates with electromagnetic radiation,''
Rev. Mod. Phys. \textbf{89}, 025001 (2017);
\href{https://doi.org/10.1103/RevModPhys.89.025001}{https://doi.org/10.1103/RevModPhys.89.025001}
[arXiv:1509.03884 [gr-qc]].

\bibitem{Bambi:2017khi}
C.~Bambi,
{\it Black Holes: A Laboratory for Testing Strong Gravity}
(Springer Singapore, 2017);
ISBN 978-981-10-4523-3, 978-981-13-5158-7, 978-981-10-4524-0;
\href{https://doi.org/10.1007/978-981-10-4524-0}{https://doi.org/10.1007/978-981-10-4524-0}

\bibitem{Bambi:2024kqz}
C.~Bambi and A.~Cardenas-Avendano,
{\it Recent Progress on Gravity Tests. Challenges and Future Perspectives}
(Springer Singapore, 2024);
ISBN 978-981--972870-1, 978-981--972873-2, 978-981--972871-8;
\href{https://doi.org/10.1007/978-981-97-2871-8}{https://doi.org/10.1007/978-981-97-2871-8}

\bibitem{Tripathi:2020yts}
A.~Tripathi, Y.~Zhang, A.~B.~Abdikamalov, D.~Ayzenberg, C.~Bambi, J.~Jiang, H.~Liu and M.~Zhou,
%``Testing General Relativity with NuSTAR data of Galactic Black Holes,''
Astrophys. J. \textbf{913}, 79 (2021);
\href{https://doi.org/10.3847/1538-4357/abf6cd}{https://doi.org/10.3847/1538-4357/abf6cd}
[arXiv:2012.10669 [astro-ph.HE]].

\bibitem{Bambi:2021chr}
C.~Bambi,
%``Testing General Relativity with black hole X-ray data: a progress report,''
Arab. J. Math. \textbf{11}, no.1, 81-90 (2022);
\href{https://doi.org/10.1007/s40065-021-00336-y}{https://doi.org/10.1007/s40065-021-00336-y}
[arXiv:2106.04084 [gr-qc]].

\bibitem{p1}
G.~Marx, 
Nature \textbf{211}, 22-23 (1966); 
\href{https://doi.org/10.1038/211022a0}{https://doi.org/10.1038/211022a0}.

\bibitem{p2}
J.~L.~Redding, 
Nature \textbf{213}, 588-589 (1967); 
\href{https://doi.org/10.1038/213588a0}{https://doi.org/10.1038/213588a0}.

\bibitem{lubin}
P.~Lubin, 
J. Br. Interplanet. Soc. (JBIS) \textbf{69}, 40-72 (2016).

\bibitem{Licquia:2014rsa}
T.~C.~Licquia and J.~A.~Newman,
%``Improved Estimates of the Milky Way{\textquoteright}s Stellar Mass and Star Formation Rate From Hierarchical Bayesian Meta-analysis,''
Astrophys. J. \textbf{806}, 96 (2015);
\href{https://doi.org/10.1088/0004-637X/806/1/96}{https://doi.org/10.1088/0004-637X/806/1/96}
[arXiv:1407.1078 [astro-ph.GA]].

\bibitem{Lutsenko:2025aal}
A.~Lutsenko, G.~Carraro, V.~Korchagin, R.~Tkachenko and K.~Vieira,
%``Counting Mass with Gaia: Mass Density of Stars and Stellar Remnants in the Solar Neighborhood,''
Astrophys. J. \textbf{990}, 88 (2025);
\href{https://doi.org/10.3847/1538-4357/adec66}{https://doi.org/10.3847/1538-4357/adec66}
[arXiv:2507.06052 [astro-ph.GA]].

\bibitem{Shvartsman}
V.~F.~Shvartsman,
Soviet Astron. AJ \textbf{15}, 377 (1971).

\bibitem{Meszaros75}
P.~Meszaros,
%{\it Radiation from spherical accretion onto black holes},
Astron. Astrophys. \textbf{44}, 59-68 (1975).

\bibitem{McDowell85}
J.~McDowell,
%{\it Accretion radiation from nearby isolated black holes},
Mon. Not. Roy. Astron. Soc. \textbf{217}, 77-85 (1985);
\href{https://doi.org/10.1093/mnras/217.1.77}{https://doi.org/10.1093/mnras/217.1.77}

\bibitem{Campana93}
S.~Campana and M.~C.~Pardi,
%{\it Do molecular clouds contain accreting black holes?},
Astron. Astrophys. \textbf{277}, 477 (1993).

\bibitem{Fujita:1997fh}
Y.~Fujita, S.~Inoue, T.~Nakamura, T.~Manmoto and K.~E.~Nakamura,
%{\it Emission from isolated black holes and MACHOs accreting from the interstellar medium},
Astrophys. J. Lett. \textbf{495}, L85 (1998);
\href{https://doi.org/10.1086/311220}{https://doi.org/10.1086/311220}
[arXiv:astro-ph/9712284 [astro-ph]].

\bibitem{Tsuna:2018abi}
D.~Tsuna, N.~Kawanaka and T.~Totani,
{\it X-ray Detectability of Accreting Isolated Black Holes in Our Galaxy},
Mon. Not. Roy. Astron. Soc. \textbf{477}, 791-801 (2018);
\href{https://doi.org/10.1093/mnras/sty699}{https://doi.org/10.1093/mnras/sty699}
[arXiv:1801.04667 [astro-ph.HE]].

\bibitem{Kimura:2021ayq}
S.~S.~Kimura, K.~Kashiyama and K.~Hotokezaka,
{\it Multiwavelength Emission from Magnetically Arrested Disks around Isolated Black Holes},
Astrophys. J. Lett. \textbf{922}, L15 (2021);
\href{https://doi.org/10.3847/2041-8213/ac35dc}{https://doi.org/10.3847/2041-8213/ac35dc}
[arXiv:2109.14389 [astro-ph.HE]].

\bibitem{Murchikova:2025oio}
L.~Murchikova and K.~Sahu,
{\it Observability of Isolated Stellar-mass Black Holes},
Astrophys. J. Lett. \textbf{988}, L12 (2025);
\href{https://doi.org/10.3847/2041-8213/ade7f8}{https://doi.org/10.3847/2041-8213/ade7f8}
[arXiv:2506.20711 [astro-ph.GA]].

\bibitem{Lubin22}
P.~Lubin,
{\it The Path to Transformational Space Exploration}
(World Scientific Publishing Company, 2022);
ISBN 978-981-12-4903-7, 978-981-12-4828-3;
\href{https://doi.org/10.1142/11918}{https://doi.org/10.1142/11918}

\bibitem{Kuhlmey25}
J.~Y.~Lin, C.~M.~de~Sterke, O.~Ilic and B.~T.~Kuhlmey,
%{\it Photonic Lightsails: Fast and Stable Propulsion for Interstellar Travel},
\href{https://doi.org/10.48550/arXiv.2502.17828}{https://doi.org/10.48550/arXiv.2502.17828}
[arXiv:2502.17828 [astro-ph.IM]].

\bibitem{Bandutunga}
C.~P.~Bandutunga, P.~G.~Sibley, M.~J.~Ireland and R.~L.~Ward,
J. Opt. Soc. Am. B \textbf{38}, 1477 (2021);
\href{https://doi.org/10.1364/JOSAB.414593}{https://doi.org/10.1364/JOSAB.414593}.

\bibitem{Mathur:2024ify}
S.~D.~Mathur and M.~Mehta,
{\it The Fuzzball Paradigm} 
in {\it The Black Hole Information Paradox: A Fifty-Year Journey}
(Eds. A.~Akil and C.~Bambi, Springer Singapore, 2025), pp 295-340;
\href{https://doi.org/10.1007/978-981-96-6170-1_11}{https://doi.org/10.1007/978-981-96-6170-1\_11}
[arXiv:2412.09495 [hep-th]].

\bibitem{Will:2014kxa}
C.~M.~Will,
%``The Confrontation between General Relativity and Experiment,''
Living Rev. Rel. \textbf{17}, 4 (2014);
\href{https://doi.org/10.12942/lrr-2014-4}{https://doi.org/10.12942/lrr-2014-4}
[arXiv:1403.7377 [gr-qc]].

\bibitem{Uzan:2002vq}
J.~P.~Uzan,
%``The Fundamental Constants and Their Variation: Observational Status and Theoretical Motivations,''
Rev. Mod. Phys. \textbf{75}, 403 (2003);
\href{https://doi.org/10.1103/RevModPhys.75.403}{https://doi.org/10.1103/RevModPhys.75.403}
[arXiv:hep-ph/0205340 [hep-ph]].

\bibitem{Uzan:2010pm}
J.~P.~Uzan,
%``Varying Constants, Gravitation and Cosmology,''
Living Rev. Rel. \textbf{14}, 2 (2011);
\href{https://doi.org/10.12942/lrr-2011-2}{https://doi.org/10.12942/lrr-2011-2}
[arXiv:1009.5514 [astro-ph.CO]].

\bibitem{Bambi:2022lhq}
C.~Bambi,
{\it Search for Variations of Fundamental Constants}
in {\it Recent Progress on Gravity Tests. Challenges and Future Perspectives}
(Eds. C.~Bambi and A.~Cardenas-Avendano, Springer Singapore, 2024), pp 417-431;
\href{https://doi.org/10.1007/978-981-97-2871-8_10}{https://doi.org/10.1007/978-981-97-2871-8\_10}
[arXiv:2210.11959 [gr-qc]].

\bibitem{Berengut:2013dta}
J.~C.~Berengut, V.~V.~Flambaum, A.~Ong, J.~K.~Webb, J.~D.~Barrow, M.~A.~Barstow, S.~P.~Preval and J.~B.~Holberg,
%``Limits on the dependence of the fine-structure constant on gravitational potential from white-dwarf spectra,''
Phys. Rev. Lett. \textbf{111}, 010801 (2013);
\href{https://doi.org/10.1103/PhysRevLett.111.010801}{https://doi.org/10.1103/PhysRevLett.111.010801}
[arXiv:1305.1337 [astro-ph.CO]].

\end{thebibliography}
\end{document}